\input psfig.sty

\def\eg{{\it e.g.}}
\def\ie{{\it i.e.}}
\def\etal{{\it et al.}}
\def\go{
\mathrel{\raise.3ex\hbox{$>$}\mkern-14mu\lower0.6ex\hbox{$\sim$}}
}   
\def\lo{
\mathrel{\raise.3ex\hbox{$<$}\mkern-14mu\lower0.6ex\hbox{$\sim$}}
}   
\def\Bf{\ifmmode B_{\phi} \else $B_{\phi}$\fi} 
\def\Bz{\ifmmode B_z \else $B_z$\fi} 
\def\rh{\ifmmode {\cal R}_{\cal H} \else ${\cal R}_{\cal H}$\fi} 
\def\Bv{\ifmmode B_{\varpi} \else $B_{\varpi}$\fi} 
\def\Cm{\ifmmode {\cal C}^{({3\over 2})}_{m-1} \else
${\cal C}^{({3\over 2})}_{m-1}$\fi} 
\def\Pm{\ifmmode {\cal P}^{(0,2)}_{m-1} \else
${\cal P}^{(0,2)}_{m-1}$\fi} 
\def\Qm{\ifmmode {\cal Q}^{(0,2)}_{m-1} \else
${\cal Q}^{(0,2)}_{m-1}$\fi} 
\def\Qabn{\ifmmode {\cal Q}^{(\alpha,\beta)}_n\else
${\cal Q}^{(\alpha,\beta)}_n$\fi} 
\def\Qabzero{\ifmmode {\cal Q}^{(\alpha,\beta)}_0\else
${\cal Q}^{(\alpha,\beta)}_0$\fi} 
\def\Qabone{\ifmmode {\cal Q}^{(\alpha,\beta)}_1\else
${\cal Q}^{(\alpha,\beta)}_1$\fi} 

\ifx\mnmacrosloaded\undefined \input mn\fi




\begintopmatter  

\title{The structure of black hole magnetospheres. 
I. Schwarzschild black holes}

\author{Pranab Ghosh$^{1, 2}$}

\affiliation{NASA Goddard Space Flight Center, Greenbelt, MD 20771, USA}
\vskip 4pt
\affiliation{$^1$Senior NAS/NRC Fellow}
\vskip 4pt
\affiliation{$^2$Permanent address: Tata Institute of Fundamental Research, 
Bombay 400 005, India, pranab@tifr.res.in}

\shortauthor{P. Ghosh}
\shorttitle{Black Hole Electrodynamics}

\abstract {We introduce a multipolar scheme for describing the structure
of stationary, axisymmetric, force-free black-hole magnetospheres in the 
``3+1'' formalism. We focus here on Schwarzschild spacetime, giving a 
complete classification of the separable solutions of the stream 
equation. We show a transparent term-by-term analogy of our solutions 
with the familiar multipoles of flat-space electrodynamics. We discuss  
electrodynamic processes around disk-fed black holes in which our 
solutions find natural applications: (a) ``interior'' solutions in studies 
of the Blandford-Znajek process of extracting the hole's rotational energy,
and of the formation of relativistic jets in active galactic 
nuclei and ``microquasars'', and, (b) ``exterior'' solutions in studies of 
accretion disk dynamos, disk-driven winds and jets. On the strength 
of existing numerical studies, we argue that the poloidal field structures 
found here are also expected to hold with good accuracy for rotating black 
holes, except for maximum possible rotation rates. We show that the 
closed-loop exterior solutions found here are not in contradiction with 
the Macdonald-Thorne theorem, since these solutions, which diverge 
logarithmically on the hole's horizon $\cal H$, apply only to those 
regions which exclude $\cal H$.}

\keywords {black hole physics -- accretion disks -- galaxies: active 
-- galaxies: nuclei -- galaxies: jets}

\maketitle  

\section{Introduction}

The study of black hole magnetospheres derives its impetus from
our desire to understand the central powerhouse in active galactic 
nuclei (AGN) and double radio sources. In a seminal paper, Blandford 
and Znajek (1977, henceforth BZ) proposed that magnetic fields 
threading a Kerr black hole's horizon $\cal H$ can tap the rotational 
energy of the hole, transporting the energy in twin beams to 
power luminous double radio lobes. The structure of force-free 
magnetospheres surrounding Kerr black holes was discussed by BZ in a 
formal four-dimensional language, giving two specific examples of 
simple field structure: (a) split monopole and (b) non-separable 
paraboloid (see \S2).

In a pioneering effort to bridge the gap between the language of 
black-hole elctrodynamics and that of flat-space elctrodynamics 
with which astrophysicists are normally familiar, 
Thorne and Macdonald (1982, henceforth TM) and Macdonald and Thorne 
(1982, henceforth MT) introduced the ``3+1'' formalism, wherein the 
spacetime was split up into three space directions and one, uniquely 
chosen, ``universal'' time direction (see \S2 for a brief sketch; a 
lucid exposition of the formalism is given in Thorne, Price \& 
Macdonald 1986, henceforth TPM). Expressed in this way, equations 
of black-hole electrodynamics become remarkably similar to those of 
 pulsar electrodynamics, and one can justifiably expect to carry 
the accomplishment, expertise, and intuition in pulsar elctrodynamics 
over to black-hole electrodynamics.

It is surprising, therefore, that a formalism for describing the 
general field structure of black-hole magnetospheres has not been
constructed so far, although the methods of the ``3+1'' formalism
are ideally suited for this. Indeed, it seems that only four specific,
simple field structures in Schwarzschild spacetime have ever been 
reported in the literature, which appear to have been  found 
{\it ad hoc}. Two of these are separable solutions of the stream 
equation (see \S2): the monopole (and its trivial 
generalization, the ``split monopole''; BZ; Macdonald 1984, henceforth 
M84) and the uniform field (Wald 1974; Hanni \& Ruffini 1976; M84). 
The others are non-separable solutions: (a) asymptotically
paraboloidal field lines (BZ), and, (b) asymptotically vertical field
lines with specified slopes on the equatorial plane (Ghosh \& 
Abramowicz 1997, henceforth GA). Apparently, the curved-spacetime
analogue of our all-too familiar and highy useful multipole 
description of electric and magnetic fields in flat-space 
electrodynamics has remained unnoticed so far.      

This paper introduces the first steps towards evolving a general 
scheme for describing black-hole magnetospheres in terms of their
basic constituents, the multipolar ``interior'' and ``exterior'' 
solutions in the appropriate spacetime. We focus in this work on 
the field structure of stationary, axisymmetric, force-free 
magnetospheres around Schwarzschild black holes. We give a complete 
characterization of separable solutions for the case in which the 
poloidal and toroidal fields are decoupled. We show that the angular 
parts of the stream functions which describe the magnetic field (see
\S2) are described by Gegenbauer polynomials. The radial parts of
the stream functions divide themselves into two classes, 
interior and exterior solutions, in analogy with the nomenclature
of flat-space electrodynamics. The former class is described by
Jacobi polynomials, and the latter, by Jacobi functions of the 
second kind. We demonstrate a clear analogy between the basic
constituents of the Schwarzschild solutions and those of flat-space
electrodynamics, identifying Schwarzschild monopoles, dipoles,
quadrupoles, and so on. We show that the existence of exterior 
solutions with closed field loops is not in conflict with a theorem 
proved by MT on the impossibility of closed field loops threading 
the horizon $\cal H$, since these solutions, which diverge 
logarithmically on $\cal H$, are appplicable only to regions which 
exclude $\cal H$, entirely in analogy with the behaviour of exterior 
solutions of flat-space electrodynamics with respect to the origin. 
Indeed, our inability to find any solutions which have closed loops 
and remain finite on $\cal H$ is a confirmation of the MT theorem.

Those studies of astrophysical processes involving black-hole
electrodynamics which benefit particularly from this 
approach include (a) BZ-process studies with ``realistic'' field 
configurations suggested by recent numerical simulations (GA and
references therein; particularly see Balbus \& Hawley 1998; 
Brandenburg \etal~1995), (b) studies of magnetic coupling of our 
exterior solutions to accretion disks in relation to dynamo action, 
jets and hydromagnetic winds (K\"onigl \& Kartje 1994; Khanna \& 
Camenzind 1992,1996; Blandford 1998), and, (c) studies of our
interior solutions in relation to magnetohydrodynamic (MHD) models 
of jet formation and collimation in AGN and ``microquasars'' 
(Camenzind 1995; Fendt 1997, henceforth F97; Koide, Shibata 
\& Kudo 1998, 1999, henceforth K98, K99; Eikenberry \etal~1998). 
Distinct field geometries relevant for these different processes 
usually emerge naturally as low-order (see \S 5.4) examples of 
different classes of solutions identified in this work: 
we indicate these briefly, deferring detailed astrophysical 
applications to later works. Indeed, a main virtue of our 
multipolar approach to black-hole elctrodynamics is that a specific 
model field-configuration is immediately identifiable as the 
natural choice for a given astrophysical problem from symmetry 
considerations alone, without detailed computations.    

Calculations for rotating black holes will be given in a
later work. We note here that, in a numerical study utilizing 
relaxation methods, M84 spun up
Schwarzschild solutions corresponding to the first three of the
four simple cases listed above, and showed that the effect of 
rotation on the {\it poloidal\/} field structure was very small 
in all cases, being barely pereceptible for $0 \le {a\over M} 
\le 0.75$, where ${a\over M}$ is the hole's angular momentum per 
unit mass. If this result is generally true, it is of basic 
importance, since it would imply that, as far as the poloidal field 
structure is concerned, the hole's rotation is not just an inessential 
complication, but also an insignificant one except at the fastest 
possible rotation rates. For astrophysical applications, the 
poloidal structures classified in this paper thus appear to be
adequate in all cases not involving black holes rotating near 
the limiting rate. Recent numerical simulations of jet formation by 
Koide and collaborators (K98; K99; Koide, Meier, Shibata \& Kudo 
1999, henceforth KMSK), using their general relativistic MHD code, 
support this conclusion: we return to this point in \S 5.4. 
In a recent numerical study of the stream equation utilizing 
finite-element techniques, F97 computed field structures 
(also see \S 5.4) in Kerr spacetime which join on smoothly to the 
asymptotic jet solutions given by Appl \& Camenzind (1993). Finally,   
it must be clear on general grounds that the {\it toroidal\/} field 
strength is expected to depend significantly on the hole's rotation 
rate. We discuss these points again in \S 5.4, suggesting that the
effects of the hole's rotation on magnetospheric structure may be of
secondary importance in studies of the BZ process and related issues, 
while they may be crucial in understanding collimation and stability 
of jets in AGN and microquasars.   

\section{Black hole electrodynamics}

\subsection{ The ``3+1'' formulation}

Only the essentials of the ``3+1'' formulation of curved spacetime
elctrodynamics are given here, since detailed expositions may be found 
in TM, MT, and TPM. The four-dimensional formulation of 
black hole electrodynamics can be reexpressed in terms of
the familiar electric, magnetic, and current three-vectors and the charge
density scalar of flat-space electrodynamics through the use of the
following prescription. At each spacetime event, one chooses a fiducial
reference frame by doing a ``3+1'' split-up of spacetime into three
space directions and one, uniquely chosen, ``universal'' time direction. 
One then decomposes the electromagnetic field tensor into the electric 
and magnetic field vectors in the usual manner, as also the four-current 
density into the current three-vector and the charge density, rewriting 
Maxwell's equations, the continuity equation for charge and current 
density, and the momentum equation in terms of these variables.    
        
In stationary black-hole electrodynamics, the natural choice for the 
fiducial frame is that of a so-called ZAMO (zero angular momentum observer; 
see Bardeen \etal~1973), who is at rest in the hole's stationary 
gravitational field, which, for rotating holes, implies further that 
the observer's angular velocity must be such that the observer's world 
lines appear orthogonal to a family of three-dimensional hypersurfaces of 
constant (Boyer-Lindquist) time (TPM). By mentally collapsing the entire 
collection of the above constant-time hypersurfaces into a single 
three-dimensional space, which we can regard as a sort of ``absolute space'',
we can envisage electrodynamics (and other physics) as occurring in this 
absolute three-space, the universal time becoming a parameter which demarks 
the evolution of matter and fields in this absolute space. This is the 
absolute-space/universal-time picturization of the 3+1 formalism, whose
analogy with Galilean relativity facilitates extensions of astrophysical 
intuition into black hole electrodynamics (TM). 

\subsection{The stream equation}

In the above framework, the structure of a stationary,
axisymmetric, force-free black hole magnetosphere is specified by a 
scalar function, $\psi$, of position, called the {\it stream function\/},
and two functions of $\psi$, namely the currrent potential, $I$, and the
angular velocity, $\Omega_F$, of the magnetic field lines (M84). The 
poloidal and toroidal components of the magnetic field are given by       
$${\bf B}_p = {\nabla\psi \times \hat{\bf\phi} \over 2\pi\varpi}
,\eqno\stepeq$$ 
and
$${\bf B}_T = - {2I \over \aleph\varpi}\hat{\bf\phi},\eqno\stepeq$$ 
respectively, with analogous expressions for electric fields,
the toroidal electric field being zero in an axisymmetric system. 
Here, $\nabla$ is the gradient operator of the absolute three-space 
referred to above, \ie, the ``space-space'' part of the  
four-dimensional metric of the spacetime that we began with (M84). 
Further, $\aleph$ is the so-called lapse function or gravitational
redshift factor, \ie, the rate of lapse of the ZAMO's proper time 
relative to that of the universal time referred to above: it is the 
``time-time'' part of the four-dimensional metric. Finally, $\varpi = 
r\sin\theta$ is the cylindrical radius, and $\hat{\bf\phi}$ is a unit 
vector in the azimuthal direction.

The name ``stream function'' was (apparently) coined by Newtonian 
analogy, since the poloidal magnetic field points everywhere along 
streamlines of $\psi$ (\ie, poloidal lines of constant $\psi$): thus,
{\it poloidal field lines are contours of constant\/} $\psi$ [see eq.(1)]. 
Uses of this description for magnetic fields are well-known in 
flat-space electrodynamics : see, \eg, Michel's (1973, henceforth M73) 
work on pulsar electrodynamics, in which $\psi$ was called a 
field-line label. Further analogies with flat-space elctrodynamics
may be made: $\psi$ is proportional to the toroidal component 
of the vector potential (which, in turn, is related to the magnetic 
flux), and $I(r_0,\theta_0)$ is the total current through the 
azimuthal loop ($r = r_0$, $\theta = \theta_0$). The stream function 
satisfies a second-order elliptic differential equation called the
stream equation (MT), and also, sometimes, the Grad-Shafranov or 
trans-field equation (Beskin 1997, henceforth B97). The equation is  
$$\nabla\cdot\left\{{\aleph\over\varpi^2}\left[1 - {(\Omega_F - 
\omega)^2\varpi^2\over\aleph^2}\right]\nabla\psi\right\} + 
{\Omega_F - \omega\over\aleph}{d\Omega_F\over d\psi}(\nabla\psi)^2$$ 
$$+{16\pi^2\over\aleph\varpi^2}I{dI\over d\psi} = 0.\eqno\stepeq$$ 
Here,  $\omega$ is the angular velocity of the ZAMO rest frame relative
to absolute space, \ie, the ``time-space'' part of the four-dimensional 
metric referred to above. 

For Schwarzschild black holes, $\Omega_F = 0 = \omega$, so that the 
stream equation simplifies considerably. The lapse function is given in
this case by 
$$\aleph = \sqrt{1 - {2M\over r}}\,.\eqno\stepeq$$ 
where $M$ is the hole's mass. (We shall use the natural units,
$G = 1 = c$ throughout this paper.) 
In this work, we study those black-hole magnetospheres in which the 
current potential $I(\psi)$ is of such form that the last term on the 
left-hand side of eq.(3) vanishes, so that the poloidal and 
toroidal fields are, in some sense, decoupled. (For a different choice 
of $I(\psi)$, useful for modeling cylindrically collimated jets, see 
Appl \& Camenzind 1993 and F97). Two types of (astrophysically) 
important situations in which this condition holds have been
described in the literature. The first type occurs when the black hole is 
immersed in a vacuum: all analytic Schwarzschild solutions reported by 
Wald 1974, Hanni \& Ruffini 1976, BZ, and M84 fall in this category. 
The second type corresponds to the situation described by GA, 
in which $I$ = constant = $I_0$, so that 
$B_T \propto \aleph^{-1}\varpi^{-1}$ (see eq.[2]). In this case, 
the toroidal field scales asymptotically as $\Bf \sim \varpi^{-1}$ far 
from the horizon $\cal H$ and diverges, of course, as $\aleph^{-1}$ 
(MT) as $\cal H$ is approached. This asymptotic behavior is reminiscent 
of some of the Newtonian model solutions of Blandford (1976, henceforth 
B76), and is relevant for electrodynamics of accretion disk-fed black 
holes, as emphasized by GA. In fact, these two types of situations are, 
to our knowledge, the only ones in which analytic Schwarzschild 
solutions of field structures around black holes have been reported so 
far. In these circumstances, the stream equation reduces to the linear 
differential equation               
$$\nabla\cdot\left\{{\aleph\over\varpi^2}\nabla\psi\right\} 
= 0\,.\eqno\stepeq$$ 

\subsection{Separable solutions} 

With the aid of equation (4) and the explicit form of $\nabla$ 
for the Schwarzschild metric, equation (5) becomes 
$$r^2{\partial\over\partial r}\left[(1-{2M\over r}){\partial\psi
\over\partial r}\right] + \sin\theta{\partial\over\partial\theta}
\left[{1\over\sin\theta}{\partial\psi\over\partial\theta}\right]
= 0\,.\eqno\stepeq$$ 
In this work, we are principally concerned with {\it separable\/}
solutions of equation (6), which offer a rich variety of multipolar 
structure. Only two non-separable solutions of equation (6) have been 
reported in the literature (BZ, M84, GA): we refer to these
at appropriate places.

Separable solutions of equation (6) are of the form
$$\psi(r,\theta) = R(r)\Theta(\theta)\,,\eqno\stepeq$$ 
where the radial part, $R$, and the angular part, $\Theta$, satisfy
equations which are obtained by combining equations (6) and (7):
$${d\over d\theta}\left[{1\over\sin\theta}{d\Theta\over d\theta}
\right] = -A{\Theta\over\sin\theta}\,,\eqno\stepeq$$ 
and   
$${d\over dr}\left[(1-{2M\over r}){dR\over dr}\right] = 
A{R\over r^2}\,.\eqno\stepeq$$ 
Here, $A$ is the separation constant.

\section{Solutions of the stream equation}

\subsection{Lowest-order solutions}

Solutions of equations (8) and (9) in the special case $A = 0$ 
constitute the lowest-order separable solutions of the stream 
equation. In the next subsection, we quantify the idea of order in 
terms of the value of $A$. 
 
The lowest-order solutions are obtained immediately on setting
$A = 0$ in eqs. (8) and (9). These are
$$\Theta = a\cos\theta + b\,,\eqno\stepeq$$ 
and
$$R = c[r + 2M\ln(r - 2M)] + d\,.\eqno\stepeq$$ 
Here, $a$, $b$, $c$, $d$ are constants. The
nature of the solutions is discussed in detail in \S4. Note that 
the special case $c = 0$, the Schwarzschild monopole, was one of 
the first solutions to be found. Note further that the special 
case $a + b = 0,\,d = 0$ is that of the separable Schwarzschild 
paraboloid, the non-separable analogue of which has been reported 
(BZ; M84) and is discussed later. Finally, note the logarithmic 
singularity at $r = 2M$ in eq.(11), which is discussed later.

\subsection{Solutions of general order}

We now give a complete characterization of general solutions of
equations (8) and (9), to be labelled by the ordinal number, $m$,
which is defined in terms of the separation constant as
$$A = m(m+1)\,.\eqno\stepeq$$ 
With the aid of equation (12), the equation for the angular 
function, $\Theta$, becomes
$$(1 - x^2){d^2\Theta\over dx^2} + m(m+1)\Theta = 0
\,.\eqno\stepeq$$ 
Here, $x = \cos\theta$. Similarly, the equation for the radial 
function, $R$, becomes
$$(1 - z^2){d^2R\over dz^2} - 2{dR\over dz} + m(m+1)R = 0
\,.\eqno\stepeq$$ 
Here, $z = {r\over M} - 1$.

\subsection{The angular function}

Solutions of equation (13) are closely related to the Gegenbauer 
polynomials (see,\eg, Szeg\"o 1939, henceforth SZ,  Abramowitz 
\& Stegun 1972, henceforth AS; Gradshteyn \& Ryzhik 1990, 
henceforth GR), the angular function of order $m$ being given by 
$$\Theta_m(x) = (1 - x^2)\Cm(x)\,,\eqno\stepeq$$ 
where \Cm~are Gegenbauer plynomials. Explicit forms for \Cm~can
be obtained with the aid of recursion relations, starting with 
expressions for ${\cal C}^{({3\over 2})}_0$ and ${\cal C}^
{({3\over 2})}_1$ (SZ). We list below the first few. Note that we 
shall henceforth suppress the superscript in \Cm, since its 
value is the same throughout this work. 
$$m = 1 \quad {\cal C}_0(x) = 1\,,$$
$$m = 2 \quad {\cal C}_1(z) = 3x\,,$$
$$m = 3 \quad {\cal C}_2(z) = {15\over 2}x^2 - {3\over 2} 
\,.\eqno\stepeq$$ 
Detailed angular structure of the solutions is discussed in
\S4. Note that the $m = 1$ solution has dipolar symmetry, 
$\Theta_1\sim\sin^2\theta$, the $m = 2$ solution has
quadrupolar symmetry, $\Theta_2\sim\sin^2\theta\cos\theta$, and
so on. This is in keeping with our experience in flat-space 
electrodynamics.  

\subsection{The radial function} 

Solutions of equation (14), which are closely related to the Jacobi 
polynomials and Jacobi functions of the second kind (see SZ for an 
excellent review), are given by (AS, GR, SZ)
$$R_m(z) \sim (1+z)^2\cases{$\Pm(z)$ &\cr $\Qm(z)$ &\cr} 
\,,\eqno\stepeq$$ 
where \Pm are Jacobi polynomials and \Qm are Jacobi functions of the 
second kind. The radial function of order $m$ is therefore given by
$$R_m(r) = r^2\cases{$\Pm$({r\over M} - 1) &\cr $\Qm$({r\over M} - 1) 
&\cr}\,.\eqno\stepeq$$ 
The Jacobi polynomials and the Jacobi functions of the second kind 
represent two distinct classes of radial solutions, whose physical 
natures are evident from their asymptotic behaviors at 
large $r$. For $r \gg M$, their scalings are (AS, SZ)
$$\Pm \sim r^{m-1}\quad,\quad \Qm \sim r^{-m-2}\,,\eqno\stepeq$$ 
so that the radial function scales as
$$R_m(r) = \cases{r^{m+1}&\cr r^{-m}&\cr}\eqno\stepeq$$ 
in the two cases, which is immediately reminiscent of the behaviors 
of the ``interior'' and ``exterior'' solutions of flat-space 
elctrodynamics. Indeed, since $r \gg M$ is the flat-space limit of
Schwarzschild results, the analogy is quite exact. Therefore, we  
henceforth refer to the solutions bearing the Jacobi polynomials as 
the interior solutions and those bearing the Jacobi functions of the 
second kind as the exterior solutions. Since the two types of solutions 
have different physical significances and mathematical properties, we 
treat them separately.       

\subsubsection{Interior solutions} 

Explicit forms for Jacobi polynomials can be obtained from the 
above references: we list below the first few. Note that henceforth we 
suppress the two superscripts in \Pm, as their values are the same 
throughout this work. 
$$m = 1 \quad {\cal P}_0(z) = 1\,,$$
$$m = 2 \quad {\cal P}_1(z) = 2z - 1\,,$$
$$m = 3 \quad {\cal P}_2(z) = {15\over 4}z^2 - {5\over 2}z - 
{1\over 4}\,.\eqno\stepeq$$ 
For our purposes, it is useful to scale the radial function 
$R_m(r)$ in terms of its asymptotic value given in equation (20).
We combine eqs.(17) and (21), and obtain
$$R_m(r) = r^{m+1}{\cal G}_m({M\over r})\eqno\stepeq$$ 
where ${\cal G}_m({M\over r})$ is a polynomial which describes the 
general-relativistic effects. The first few of these polynomials
are
$$m = 1 \quad {\cal G}_1(x) = 1\,,$$
$$m = 2 \quad {\cal G}_2(x) = 1 - {3\over 2}x\,,$$
$$m = 3 \quad {\cal G}_3(x) = 1 - {8\over 3}x + {8\over 5}x^2
\,.\eqno\stepeq$$ 
The structure of the interior solutions is discussed in \S4. 

\subsubsection{Exterior solutions}

As the explicit forms for Jacobi functions of the second kind are not
easily found in the standard references, we describe them here at the
level of detail adequate for our purposes. We introduce a prescription
for obtaining a general function of this type, \Qabn, which, to our
knowledge, is not found in the standard mathematical references. We
suggest that explicit forms for \Qabn can be found most conveniently 
by the following prescription, through the use of recursion 
relations, starting with a suitable (integral or differential) 
representation for \Qabzero.

For arbitrary values of $\alpha$, $\beta$ satisfying the constraint
$\alpha + \beta + 1 > 0$, a useful integral representation (SZ)
for \Qabzero~is    
$$\Qabzero(z) = -2^{\alpha + \beta}{\Gamma(\alpha + 1)\Gamma(\beta + 1)
\over\Gamma(\alpha + \beta +1)}\cdot$$
$$\int_{\infty}^z{dt\over (t-1)^{\alpha + 1}(t+1)^{\beta + 1}}
\,.\eqno\stepeq$$ 
For positive integral values of $\alpha$ and $\beta$, as in our case, 
a differential representation, which leads to much faster calculation 
of explicit expressions for \Qabzero, can be obtained from equation
(24). The representation is
$$\Qabzero(z) = {(-)^{\beta + 1}2^{\alpha + \beta}\over\Gamma(\alpha + 
\beta +1)}\cdot$$
$$\left[{d^{\alpha}\over da^{\alpha}}{d^{\beta}\over db^{\beta}}
\left\{{1\over a+b}\ln{z-a\over z+b}\right\}\right]_{a = b = 1}
\,,\eqno\stepeq$$ 
and we have {\it not\/} found it in any standard reference. To prove 
equation (25), consider the identity 
$$\int_{\infty}^z{dt\over (t-a)(t+b)} = {1\over a+b}\ln{z-a\over z+b}
\,.\eqno\stepeq$$ 
Differentiation of both sides of equation (26) $\alpha$ times with 
respect to $a$ and $\beta$ times with respect to $b$, and comparison 
with equation (24), leads to equation (25).        

Recursion relations from which \Qabn~with $n > 0$ can be obtained are 
as follows (SZ). The relation used for calculating \Qabone~is 
a special one:
$$\Qabone(z) = {1\over 2}[(\alpha + \beta + 2)z + \alpha - \beta]
\Qabzero(z) - 2^{\alpha + \beta - 1}\cdot$$
$${\alpha + \beta +2\over (z-1)^{\alpha}(z+1)^{\beta}}{\Gamma(\alpha
 + 1)\Gamma(\beta + 1)\over\Gamma(\alpha + \beta +2)}
\,.\eqno\stepeq$$ 
For $n\ge 2$, \Qabn~are obtained by repeated application of the general
relation which holds for $n\ge 2$, namely 
$$2n(n + \alpha + \beta)(2n + \alpha + \beta - 2)\Qabn(z) =$$
$$(2n + \alpha + \beta - 1)[(2n + \alpha + \beta)(2n + \alpha + 
\beta - 2)z + \alpha^2 - \beta^2]{\cal Q}^{(\alpha,\beta)}_{n-1}(z)$$
$$-2(n + \alpha - 1)(n + \beta - 1)(2n + \alpha + \beta){\cal Q}^
{(\alpha,\beta)}_{n-2}(z)\,.\eqno\stepeq$$ 

We now set $\alpha = 0$ and $\beta = 2$ for our problem, and
calculate Jacobi functions of interest from equations (25), (27) 
and (28). We give the first few below, henceforth suppressing 
the two superscripts in \Qm, since their values remain the same 
throughout this work:
$$m = 1 \quad {\cal Q}_0(z) = {1\over 2}\ln{z+1\over z-1} - 
{z+2\over(z+1)^2}\,,$$

$$m = 2 \quad {\cal Q}_1(z) = (z - {1\over 2})\ln{z+1\over z-1} -  
{2z^2+3z+{2\over 3}\over(z+1)^2}\,,$$

$$m = 3 \quad {\cal Q}_2(z) =  ({15\over 8}z^2 - {5\over 4}z - 
{1\over 8})\ln{z+1\over z-1}$$
$$ - {{15\over 4}z^3 + 5z^2 - {z\over 4} - {4\over3}\over(z+1)^2}
\,.\eqno\stepeq$$ 

The structure of the exterior solutions is discussed in \S4. Note 
the appearance of a logarithmic singularity, $\sim\ln
(r-2M)$ ($z$ in eqs.[29] is given by $z = {r\over M} - 1$; see above), 
in these solutions at the horizon $\cal H$; the singularity also 
appeared in one of the lowest-order solutions discussed above. This 
singularity is reminiscent of the one which occurs in the well-known 
flat-space exterior solutions ($\sim r^{-m}$) at $r = 0$, but it is 
weaker. Of course, such a singularity does not imply the invalidity 
of exterior solutions for black hole electrodynamics, any more than 
the above power-law singularity implies the invalidity of exterior 
solutions in flat-space electrodynamics: it simply means that these 
exterior solutions are valid in regions of space which exclude 
$\cal H$, just as the regions of validity of flat-space exterior 
solutions necessarily exclude the origin. We come back to this  
point in \S4 and \S5. Finally, we note that the power-law 
singularity displayed by the Jacobi functions of equation (29) at 
$r = 0$ (\ie, $z+1=0$) has no physical significance, since these 
solutions do not apply for $r < 2M$.        

\section{Structure of the solutions}

\subsection{Lowest-order solutions}

The solution corresponding to $c = 0$ in equation (11), \ie,
$$\psi\sim a\cos\theta + b\,,\eqno\stepeq$$ 
with $a$ and $b$ constant, is the Schwarzschild monopole, 
the field lines of which are shown in Figure 1.

\beginfigure{1}
\vskip 5mm
\caption{{\bf Figure 1.} Poloidal field lines of the Schwarzschild
monopole. The horizon $\cal H$ is shown by the thick line. All
lengths are in natural units, as indicated.}
\endfigure

This solution has been discussed previously by BZ and M84, and its 
flat-space version (identical in form to eq.[30]) by M73. 
Imposition of the solenoidal condition on ${\bf B}$, which must be
satisfied in all astrophysical applications, is achieved by the  
device of assigning opposite polarities to the radial magnetic field 
in the two hemispheres (BZ, M84), \ie, a so-called ``split monopole'' 
configuration, maintained by a toroidal current in the accretion disk 
lying in the equatorial plane. Results of spinning up the split monopole 
field have been described in various approximations by BZ and M84, and 
their flat-space analogue is an exact solution given by M73. 

The solution corresponding to $a + b = 0,\,d = 0$, \ie, 
$$\psi\sim [r + 2\ln(r-2M)][1 - \cos\theta]\,,\eqno\stepeq$$ 
represents the separable Schwarzschild paraboloid (our nomenclature
comes from the observation that, in eq.[31], the sole difference from 
a familiar flat-space paraboloid is the presence of a logarithmic 
term characteristic of Schwarzschild spacetime), the field lines of 
which are shown in Figure 2. 

\beginfigure{2}
\vskip 5mm
\caption{{\bf Figure 2.} Same as Figure 1, but for the separable
Schwarzschild paraboloid.}
\endfigure

At sufficiently large radii, this configuration reduces to the 
paraboloidal field-lines of flat-space electrodynamics, which have been 
studied at length by B76. At small radii, the separable Schwarzschild 
paraboloid has one curious relativistic feature: close to $\cal H$, the 
field lines change direction, so that $\psi$, which is a measure of the 
magnetic flux (see \S2), changes sign. For the specific stream function 
given in equation (31), $\psi$ changes sign at $r\approx 2.314M$. 
Note that this asymptotically-paraboloidal separable solution has 
similarities to the corresponding non-separable solution (BZ, M84), 
which we discuss in \S5, but there are also clear differences.

\subsection{General solutions}

We now combine the angular and radial parts of general solutions for 
the stream function described in \S3, and discuss the structure of the 
resultant field configurations.

\subsubsection{Interior solutions}

The stream function for $m = 1$, namely,
$$\psi_1\sim r^2\sin^2\theta\,,\eqno\stepeq$$ 
represents a uniform magnetic field pointing in the $z$-direction,
as shown in Figure 3. 

\beginfigure{3}
\vskip 5mm
\caption{{\bf Figure 3.} Same as Figure 1, but for the uniform
field in Schwarzschild spacetime.}
\endfigure

This appears to be the most thoroughly studied (Wald 1974; Hanni \& 
Ruffini 1976; M84) field configuration in Schwarzschild spacetime, 
because of its simplicity. The configuration is identical to that 
obtained in flat-space electrodynamics corresponding to the $m = 1$ 
term of a multipolar decomposition of the scalar potential, \ie, 
$\Phi_1 = rP_1(\cos\theta)$ (here $P_1$ is a Legendre polynomial).

The stream function for $m = 2$, namely,
$$\psi_2\sim (r^3 - {3\over 2}Mr^2)\sin^2\theta\cos\theta
\,,\eqno\stepeq$$ 
represents the interior solution of the next higher order, the
field lines of which are shown in Fig.4. 

\beginfigure{4}
\vskip 5mm
\caption{{\bf Figure 4.} Same as Figure 1, but for the $m = 2$
interior solution.}
\endfigure

This configuration is very similar, but not identical, to that 
obtained in flat-space electrodynamics corresponding to the $m = 2$ 
term of a multipolar decomposition of the scalar potential, \ie, 
$\Phi_2 = r^2P_2(\cos\theta)$.

The structure of higher-order solutions may be readily obtained
in a similar manner. It is this term-by-term analogy with 
well-known configurations in flat-space electrodynamics that makes
our scheme of classifying the field-configurations in curved 
spacetime particularly transparent, and underscores the power of 
the 3+1 formalism.     

\subsubsection{Exterior solutions}

The stream function for $m = 1$, namely,
$$\psi_1\sim \left[{r^2\over 2}\ln{r\over r - 2M} - rM - M^2
\right]\sin^2\theta\,,\eqno\stepeq$$ 
represents a solution which we name the Schwarzschild dipole. The
field lines, shown in Figure 5, readily explain the terminology,
demonstrating that this solution is the Schwarzschild analogue of 
the familiar flat-space dipole field. 

\beginfigure{5}
\vskip 5mm
\caption{{\bf Figure 5.} Same as Figure 1, but for the 
Schwarzschild dipole.}
\endfigure

The analogy is quite exact since, on taking the limit $r \gg M$ in
eq.(34), we find that $\psi_1\sim{1\over r}\sin^2\theta$, which
describes the field lines of a flat-space dipole.  

Similarly, the stream function for $m = 2$, namely,
$$\psi_2\sim[(r^3 - {3\over 2}Mr^2)\ln{r\over r - 2M} - 2r^2M
+ rM^2 +$$
$${M^3\over 3}]\sin^2\theta\cos\theta\,,\eqno\stepeq$$ 
represents the Schwarzschild quadrupole, the field lines of which 
are displayed in Figure 6. 

\beginfigure{6}
\vskip 5mm
\caption{{\bf Figure 6.} Same as Figure 1, but for Schwarzschild 
quadrupole.}
\endfigure

The quadrupolar symmetry is evident. The reader can verify the 
exactness of the flat-space analogy by extracting the $r \gg M$ limit 
of equation (35), and noting that, in this limit, $\psi_2 = $ const 
is precisely the equation for the field lines of a flat-space 
quadrupole.   

The higher exterior multipoles (octupole and so on) may be readily 
obtained in a similar manner. For the exterior solutions, the 
exactness of the analogy with the multipoles of flat-space
electrodynamics makes the classification scheme completetly natural
and transparent, and re-emphasizes the virtue of the 3+1 formalism. 

\section{Discussion}

\subsection{Astrophysical context}

The classes of solutions detailed in \S\S 3-4 provide natural generic
choices for appropriate classes of black-hole electrodynamics problems 
involving accretion disks. For studies of field structure and transport
of energy or angular momentum close to the hole's horizon, our 
lowest-order solutions and interior solutions (BZ; M84; B97; KMSK) 
are clearly the best choice (also see \S 5.4). 
The uniform-field configuration, corresponding to $m=1$, describes 
the essential astrophysics very close to $\cal H$, as has been 
intuitively realized long ago (see TPM, particularly their Fig.36). 
This is a result of the physical mechanisms 
described in MT and TPM by which the black hole ``cleans'' the magnetic 
field threading its horizon of details, kinks, and closed loops (also
see \S5.3). At slightly larger distances from $\cal H$ ($\sim\rh$, say,
where \rh~is the characteristic scale size of $\cal H$), further 
details of field structure begin to appear even under the most ideal 
circumstances, due to the contribution of the currents circulating at 
the inner edge of the accretion disk. These are described by interior 
solutions of higher order, $m=2,3,....$, whose relative contributions 
determine the shape of the beam or jet of relativistic particles 
transporting energy to the radio lobes. Similar arguments apparently 
also hold for MHD jets in AGN and quasars, although the stream equation     
there is somewhat different (B97 and references therein), and so is the
structure of the interior solutions.   

On the other hand, for describing the field structure on and around the 
accretion disk at larger distances ($\gg\rh$) from $\cal H$, our 
exterior solutions are the natural choice. They describe, term by term,
the multipole moments of the currents circulating in the disk, seen as
an external observer comes closer and closer to the hole. The $m=1$
dipole is thus a first estimate of the total effect of the disk 
currents, as seen by a distant observer. This decomposition proves 
particularly useful for analyzing the relative effects of currents 
circulating in the disk and on the hole's horizon. Astrophysical 
applications of these field structures to disk-driven jets and MHD
winds, and to dynamo-action in disks are well-known (Konigl \& Kartje 
1994; Khanna \& Camenzind 1992,1996; Camenzind 1995).

Construction of model global field-structures relevant to almost all
black-hole-accretion-disk problems involves the matching of interior and 
exterior solutions at a suitable intermediate radius according to a 
prescription appropriate for the particular problem. This has direct
relevance to the ``impedance matching'' considerations long known to
be of crucial for studies of BZ processes and related particle 
acceleration issues (Lovelace, MacAuslan \& Burns 1979; MT; TPM). 
In fact, techniques borrowed from matching procedures in 
flat-space electrodynamics may provide the most rigorous
formulation of the idea of impedance matching between the black hole
source and the distant astrophysical load: we shall describe this
elsewhere.

\subsection{Non-separable solutions}

Separable solutions do not exhaust all possibilities for solutions
of equation (5). One particular non-separable solution of astrophysical
interest is that found by BZ and discussed by M84. It is given by
$$\psi\sim[(r-2M)(1-\cos\theta) - 2M(1+\cos\theta)
\ln(1+\cos\theta)]\,,\eqno\stepeq$$ 
and its field lines are asymptotically paraboloidal. This solution was
used in the original BZ study of the extraction of rotational energy 
from black holes. Equation (37) is one of the lowest-order, 
non-separable solutions of equation (5), analogous to the lowest-order 
separable paraboloid solution described by equation (31).
At large radii, both of these solutions reduce to the familiar,
exact paraboloids of flat-space electrodynamics, $\psi_{nr}\sim 
r(1-\cos\theta)$, astrophysical applications of which to 
accretion-disk electrodynamics and jet-formation have been discussed 
at length by B76.

However, there is nothing fundamental about the paraboloidal shape in 
the astrophysical context, except that field lines do need to be
asymptotically vertical if electromagnetic energy is to be beamed
along the black hole's rotation axis, as in the BZ process. In fact,  
the asymptotically paraboloidal shapes discussed above have the 
disadvantage that they have an inflexible shape on the equatorial plane 
(the ratio \Bv/\Bz~is always unity on this plane in the Newtonian limit), 
so that it is sometimes impossible to satisfy boundary conditions of
astrophysical interest on the accretion-disk surface. A good example 
is discussed in GA: recent numerical simulations of dynamo-generated 
magnetic fields in accretion disks suggest \Bv/\Bz$\sim$ 2 to 3 (GA and 
references therein), which paraboloidal field lines cannot mimic. To
achieve this, GA found a solution  
$$\psi\sim[(r-2M)(1-\cos\theta) - r_0\cos\theta$$ 
$$- 2M(1+\cos\theta)\ln(1+\cos\theta)]\,,\eqno\stepeq$$ 
which was adequate for this kind of modelling. It is clear that the
solution given by equation (37) is a linear superposition of those
given by eqs. (36) and (30), and therefore must necessarily be a 
solution of the linear equation (5). Equations (36) and (37) appear 
to be the only astrophysically interesting non-separable solutions 
of equation (5) known at this time.

There is no difficulty in obtaining further non-separable solutions 
by superposing equation (36) with other classes of separable solutions 
found in this work, but this is not of basic importance. A systematic 
method of classifying higher-order non-separable solutions would be 
more important both in principle and in the astrophysical context: 
this will be discussed elsewhere. 

\subsection{The MT theorem}

MT proved an intersting theorem (henceforth referred to as the 
MT theorem) which states that closed loops of magnetic field threading 
the horizon $\cal H$ cannot exist in stationary, axisymmetric, 
force-free magnetospheres. Therefore, a natural question which may 
arise at this point is: how do the external solutions found here, all 
of which have closed loops extending to $\cal H$, stand {\it vis \`a
vis\/} the MT theorem?

To answer this question, consider first the proof of the MT theorem.
Since we are concerned here only with Schwarzschild black holes, we
shall be content with the proof for this case, which is
particularly simple. Since equation (5) implies that the vector
${\aleph\over\varpi^2}\nabla\psi$ is solenoidal, we can convert it
, by Gauss's theorem, into the condition for vanishing of its 
surface integral over a closed surface $\cal L$ which consists of 
a closed loop and the region between its footpoints on $\cal H$:
$$\int_{\cal L}{\aleph\over\varpi^2}\nabla\psi\cdot d{\bf S}
= 0\,.\eqno\stepeq$$ 
Here, $d{\bf S}$ is an element of surface area, its direction 
being along the outward normal. 
Now, MT considered stream functions which (and whose derivatives) 
remained finite on $\cal H$, so that their next argument was that,
since the lapse function $\aleph$ vanishes on $\cal H$ by 
definition (see, \eg, Bardeen 1973), so does that part of the 
integral which comes from $\cal H$, implying that the integral
over the closed field loop itself vanishes. But, since the field
lines are contours of constant $\psi$ (see \S2.2), $\nabla\psi$
is everywhere parallel to $d{\bf S}$, so that the integrand has the
same sign all over the closed loop. It follows, therefore, that
the integral can vanish only if the integrand vanishes everywhere,
which means that there can be no field or field loop.

Consider now the results of applying the MT theorem to our 
closed-loop external solutions. Since the stream functions 
$\psi$ for these solutions diverge like 
$\ln\aleph$ as $\cal H$ is approached, as explained earlier 
(see \S3.4.2 and eq.[4]), $\nabla\psi$ diverges as $\aleph^{-1}$.
Thus the integrand, ${\aleph\over\varpi^2}\nabla\psi$, is finite
over $\cal H$, and so is its contribution to the integral in eq.
(38). Depending on the details of the field geometry in the close 
neighborhood of $\cal H$, it may, therefore, be possible to 
satisfy eq.(38) without requiring that the field be zero everywhere
on the closed loop. Thus, the MT theorem is not in contradiction
with closed-loop solutions that diverge on $\cal H$. The situation
is closely analogous to that for exterior solutions in flat-space 
electrodynamics: since these solutions, which have closed field 
loops, diverge at the origin ($r = 0$), integrals similar to that 
in eq.(38) (or related integrals involving the scalar potential) 
have to be evaluated carefully in the neighbourhood of the origin, 
and can have finite contributions from there.

The above discussion does not imply, of course, that the exterior 
solutions found here can extend upto $\cal H$. This is ruled out
simply because they diverge on $\cal H$ and are, therefore, 
physically inadmissible there. As we have indicated earlier, their
domain of validity is a region of space which excludes $\cal H$,
as the name exterior solution implies. Examples of this abound in
flat-space electrodynamics, and it is not difficult to envisage
model problems in black-hole electrodynamics where an inner region
containing $\cal H$ admits of interior solutions, surrounded by an
outer region which admits of exterior solutions: we have already 
indicated above that matching these solutions may be of considerable 
astrophysical importance.

Finally, the MT theorem only excludes the existence of 
closed-loop field configurations which remain finite on $\cal H$.
We have shown above that all separable Schwarzschild solutions
with closed loops diverge on $\cal H$. Since the two non-separable
solutions known so far (see above) do not have closed loops, it
is correct to say that all known Schwarzschild solutions with
closed loops diverge on $\cal H$. This is thus a {\it confirmation
\/} of the MT theorem, showing that the configurations that MT showed
to be untenable are, indeed, not found amnong the known solutions.
Again, we find here a close analogy with multipolar expansion in 
flat-space electrodynamics, where all of the closed-loop multipole
solutions (dipole, quadrupole, etc) diverge at the origin.   

\subsection{Concluding remarks}
 
Our confidence in the relevance of the poloidal Schwarzschild 
solutions, found in this work, to electrodynamics around rotating black
holes stems largely from the insignificant influence of rotation, except
at the highest possible rates, on poloidal field structures found in
all cases investigated so far (M84; K98; K99; KMSK; also see below). 
A remaining concern is about the role of the toroidal field, whose 
strength is expected to depend significantly on the hole's rotation rate 
(MT). While poloidal and toroidal fields are effectively decoupled, as we
argued in \S2, in most problems of interest in black-hole
electrodynamics, so that the higher toroidal fields generated by 
increasing rotation rates do not react back on the poloidal fields, 
their influences on the essential astrophysical phenomena, \eg, the
BZ process, need to be clarified. The key point here is the effect 
of the well-known boundary conditions on ${\cal H}$ (MT and references
therein). As we approach ${\cal H}$, the poloidal field becomes 
entirely normal, $B_{\perp}$, to ${\cal H}$, while the toroidal field
$B_{T}$ diverges as $\aleph^{-1}$, so that $B_{H}\equiv\aleph B_{T}$
remains finite. The strength of the BZ process scales with the magnetic
stress $B_{\perp}B_{H}$, but the `degeneracy' condition ($\bf{E.B}=0$)
outside $\cal H$ fixes the ratio $B_{\perp/}B_{H}$ in terms of
the hole's rotation rate (GA and references therein), so that the
BZ power output finally scales with $B_{\perp}^2$ alone. Thus, any
direct influence of the magnitude of the toroidal field on the 
strength of this particular astrophysical process ultimately drops 
out, and a sufficiently accurate estimate of the poloidal field close 
to ${\cal H}$ is all one needs to calculate the strength of the BZ 
process and related quantities for a black hole with a specified 
rotation rate. This is why our poloidal field calculations are 
expected to be of considerable practical value in BZ-process 
calculations relevant to double radio sources, AGN, and quasars, as
also to the poloidal field structures associated with relativistic
jets in AGN and microquasars. However, there are other astrophysical 
phenomena that do depend crucially on the strength of $B_{T}$ away 
from the horizon, such as the collimation and stability of the above 
jets (Appl \& Camenzind 1992; K99; KMSK). 
We expect, therefore, that adequate investigations of these phenomena 
will require reliable calculations of the toroidal field structure of
the magnetospheres of rotating black holes, which is beyond the scope 
of this work.

After the original version of this paper was submitted, we became aware
of an excellent contaporaneous review (B97) on axisymmetric stationary
flows in astrophysical objects, in which a model black-hole magnetosphere
is constructed by expanding the nonrelativistic M73 solution referred
to earlier, and introducing relativistic effects into each term 
separately. The final result, given upto two terms, seems to be similar 
to a combination of our interior solutions with $m=1,3$; the $m=2$ term 
vanishes for this particular solution considered by B97 because of 
its symmetry properties. We also note the similarity between the 
poloidal structure of our separable Schwarzschild paraboloid (Fig.2), 
and that of the numerical computation in F97 of collimated 
jet magnetospheres in Kerr spacetime (Fig. 3 of F97): this may well be 
indicative of the secondary importance of the hole's rotation on 
magnetospheric structure that we have suggested in this paper. 
Finally, the results of recent simulations of jet formation by Koide and 
collaborators (K98; K99; KMSK) with the aid of their general relativistic 
MHD code clearly support our conclusions. The poloidal field structures 
around rapidly-rotating Kerr holes (KMSK) are found by these authors to be 
very similar to those around Schwarzschild holes (K98), particularly for 
co-rotating Keplerian disks, confirming a major argument used in this 
paper. Further, these authors start their 
simulations with a uniform poloidal magnetic field (our interior 
solution with $m=1$, as shown in our Fig.3), and obtain poloidal 
structures at the end of their runs which appear remarkably similar to 
our separable Schwarzschild paraboloid (a lowest-order solution shown in 
our Fig.2), for both Schwarzschild holes (Fig.2 of K98) and 
rapidly-rotating Kerr holes (Fig.1(d) of KMSK) accreting from 
co-rotating Keplerian disks. This gives us much confidence in the 
relevance of the basic ``building block'' poloidal field structures found 
in this work to essential astrophysical processes near black holes: 
future simulations should fully exploit our multipolar expansion scheme.

\bye